\documentclass[11pt,a4paper,showpacs,showkeys,,superscriptaddress]{article}

\usepackage{epsfig}
\usepackage{amsmath,amssymb}
\usepackage{graphicx}
\usepackage{xcolor}
\usepackage{subfigure}

\usepackage{latexsym}
\usepackage{hyperref}
\hypersetup{colorlinks,%
  citecolor=blue,%
  linkcolor=red,%
  pdftex}

\begin{document}

\baselineskip 6mm
\renewcommand{\thefootnote}{\fnsymbol{footnote}}

\newcommand{\nc}{\newcommand}
\newcommand{\rnc}{\renewcommand}



\newcommand{\tcb}{\textcolor{blue}}
\newcommand{\tcr}{\textcolor{red}}
\newcommand{\tcg}{\textcolor{green}}


\def\beq{\begin{equation}}
\def\eeq{\end{equation}}
\def\ba{\begin{array}}
\def\ea{\end{array}}
\def\bea{\begin{eqnarray}}
\def\eea{\end{eqnarray}}
\def\nn{\nonumber}


\def\CMP{Commun. Math. Phys.~}
\def\JHEP{JHEP~}
\def\Pre{Preprint}
\def\PRL{Phys. Rev. Lett.~}
\def\PR {Phys. Rev.~}
\def\CQG {Class. Quant. Grav.~}
\def\PL {Phys. Lett.~}
\def\NP {Nucl. Phys.~}

\def\G{\Gamma}

\def\S{{\bf S}}
\def\C{{\bf C}}
\def\Z{{\bf Z}}
\def\R{{\bf R}}
\def\N{{\bf N}}
\def\M{{\bf M}}
\def\P{{\bf P}}
\def\bm{{\bf m}}
\def\bn{{\bf n}}

\def\CA{{\cal A}}
\def\CB{{\cal B}}
\def\CC{{\cal C}}
\def\CD{{\cal D}}
\def\CE{{\cal E}}
\def\CF{{\cal F}}
\def\CH{{\cal H}}
\def\CM{{\cal M}}
\def\CG{{\cal G}}
\def\CI{{\cal I}}
\def\CJ{{\cal J}}
\def\CL{{\cal L}}
\def\CK{{\cal K}}
\def\CN{{\cal N}}
\def\CO{{\cal O}}
\def\CP{{\cal P}}
\def\CQ{{\cal Q}}
\def\CR{{\cal R}}
\def\CS{{\cal S}}
\def\CT{{\cal T}}
\def\CU{{\cal U}}
\def\CV{{\cal V}}
\def\CW{{\cal W}}
\def\CX{{\cal X}}
\def\CY{{\cal Y}}
\def\CZ{{\cal Z}}

\def\We{{W_{\mbox{eff}}}}


\newcommand{\Lie}{\pounds}

\newcommand{\p}{\partial}
\newcommand{\bp}{\bar{\partial}}

\newcommand{\half}{\frac{1}{2}}

\newcommand{\bfalpha}{{\mbox{\boldmath $\alpha$}}}
\newcommand{\bfbeta}{{\mbox{\boldmath $\beta$}}}
\newcommand{\bfgamma}{{\mbox{\boldmath $\gamma$}}}
\newcommand{\bfmu}{{\mbox{\boldmath $\mu$}}}
\newcommand{\bfpi}{{\mbox{\boldmath $\pi$}}}
\newcommand{\bfvarpi}{{\mbox{\boldmath $\varpi$}}}
\newcommand{\bftau}{{\mbox{\boldmath $\tau$}}}
\newcommand{\bfeta}{{\mbox{\boldmath $\eta$}}}
\newcommand{\bfxi}{{\mbox{\boldmath $\xi$}}}
\newcommand{\bfkappa}{{\mbox{\boldmath $\kappa$}}}
\newcommand{\bfepsilon}{{\mbox{\boldmath $\epsilon$}}}
\newcommand{\bfTheta}{{\mbox{\boldmath $\Theta$}}}

\newcommand{\bz}{{\bar{z}}}

\newcommand{\dalpha}{\dot{\alpha}}
\newcommand{\dbeta}{\dot{\beta}}
\newcommand{\blambda}{\bar{\lambda}}
\newcommand{\btheta}{{\bar{\theta}}}
\newcommand{\bsigma}{{{\bar{\sigma}}}}
\newcommand{\bepsilon}{{\bar{\epsilon}}}
\newcommand{\bpsi}{{\bar{\psi}}}


\def\ct{\cite}
\def\la{\label}
\def\eq#1{(\ref{#1})}


\def\a{\alpha}
\def\b{\beta}
\def\g{\gamma}
\def\G{\Gamma}
\def\d{\delta}
\def\D{\Delta}
\def\ep{\epsilon}
\def\e{\eta}
\def\ph{\phi}
\def\Ph{\Phi}
\def\ps{\psi}
\def\Ps{\Psi}
\def\k{\kappa}
\def\l{\lambda}
\def\L{\Lambda}
\def\m{\mu}
\def\n{\nu}
\def\th{\theta}
\def\Th{\Theta}
\def\r{\rho}
\def\s{\sigma}
\def\S{\Sigma}
\def\ta{\tau}
\def\o{\omega}
\def\O{\Omega}
\def\pr{\prime}
\def\f{\varphi}


\def\half{\frac{1}{2}}

\def\goto{\rightarrow}

\def\na{\nabla}
\def\grad{\nabla}
\def\curl{\nabla\times}
\def\div{\nabla\cdot}
\def\pa{\partial}

\def\bra{\left\langle}
\def\ket{\right\rangle}
\def\lb{\left[}
\def\lc{\left\{}
\def\ls{\left(}
\def\lp{\left.}
\def\rp{\right.}
\def\rb{\right]}
\def\rc{\right\}}
\def\rs{\right)}
\def\cl{\mathcal{l}}

\def\vac#1{\mid #1 \rangle}

\def\td#1{\tilde{#1}}
\def\check{ \maltese {\bf Check!}}


\def\Tr{{\rm Tr}\,}
\def\det{{\rm det}\,}


\def\bc#1{\nnindent {\bf $\bullet$ #1} \\ }
\def\ch {$<Check!>$ }
\def\ss {\vspace{1.5cm}}

\begin{titlepage}
%
%
%
%
%
%
%
%
\begin{center}
{\Large \bf Scaling symmetry and scalar hairy rotating $AdS_{3}$ black holes}
%
\vskip 1. cm
  { Byoungjoon Ahn\footnote{e-mail : bjahn@yonsei.ac.kr}, Seungjoon Hyun\footnote{e-mail : sjhyun@yonsei.ac.kr}, Sang-A Park\footnote{e-mail : sangapark@yonsei.ac.kr},
  Sang-Heon Yi\footnote{e-mail : shyi@yonsei.ac.kr} 
 }
\vskip 0.5cm
{\it Department of Physics, College of Science, Yonsei University, Seoul 120-749, Korea}\\
\end{center}
\thispagestyle{empty}
\vskip1.5cm
%
%
\centerline{\bf ABSTRACT} \vskip 4mm
 \vspace{1cm} 
\noindent  
By using the  scaling symmetry in the reduced action formalism, we derive the novel Smarr relation which holds even for the hairy rotating $AdS_{3}$ black holes. And then,  by using the Smarr relation we argue that the hairy rotating $AdS_{3}$ black holes are stable thermodynamically, compared to the non-hairy ones.  
\vspace{2cm}
%
%
\end{titlepage}
\setcounter{footnote}{0}
%
%
%
%

\section{Introduction}
Anti-de Sitter(AdS) space and its related spaces become  very interesting research subjects  possessing the diverse applications after the birth of the gauge/gravity correspondence~\cite{Maldacena:1997re}. Since some physical features of the strong coupling regime in the dual field theory may be understood, at least qualitatively, through the classical computation in gravity on the AdS space, various classical results in gravity took on quite a new aspect  as the counterpart to the quantum field theory.  In this regard, black holes have much implication and their physics has been resurfaced as the theoretical tool for strong coupling dynamics in field theory. One of interesting aspects of AdS black holes      is that they allow the scalar hairy deformation, which corresponds to the deformation of the dual conformal field theory by the operator dual to the scalar field.

Though some analytic solutions for scalar hairy  $AdS_{3}$ black holes have been obtained~\cite{Henneaux:2002wm,Henneaux:2004zi}, they are rather complicated and so it is not easy to assure their stability when they are non-supersymmetric.  The  stability issue after the scalar hairy deformation would be related to  the end point of the renormalization group flow in the dual field theory deformed by the dual operator.  From the field theory perspective, the behavior under the renormalization group flow would be determined by the nature of the perturbing operator dual to the scalar and  has nothing to do with the existence of the analytic solution of AdS black holes. This line of thought seems to indicate that the stability of scalar hairy deformed black holes might be understood, regardless of the existence of the analytic solutions, only by using some properties of scalar hairy  black holes. 

There are at least two kinds of stability concepts in black hole physics. One of them is the dynamical stability understood usually  by the linear perturbation analysis around the black hole background,  which seems to require some detailed information on black hole solutions.  The other one is the thermodynamic stability whose criterion is just their entropy values.  Since the black hole entropy is universal and determined by the area law~\cite{Bekenstein:1973ur,Hawking:1974sw}, the thermodynamic stability has the chance to be determined through the less information on black hole solutions. In this paper, we would like to suggest  the Smarr relation~\cite{Smarr:1972kt} as the information on black holes, which allows us to determine the relative stability between hairy and non-hairy black holes. 

To derive the novel Smarr relation without requiring the analytic solutions, we use the scaling symmetry of the reduced action, which
 has been shown to lead to the Smarr relation for static black holes with the exact Killing vectors for mass and angular momentum~\cite{Banados:2005hm,Hyun:2015tia}. In this paper, we extend this symmetry to the hairy rotating AdS black holes which allow  a single Killing vector only. Though  our analysis is performed in the case of the three-dimensional Einstein gravity, it can be applied straightforwardly to higher dimensional planar black holes and  to higher derivative gravity. We  give some comments on the mass expression on the scalar hairy AdS black holes. 
 
\section{Rotating hairy black holes}
In this section, we  give the model of three-dimensional rotating AdS black holes with a scalar hair.   The action of three-dimensional Einstein gravity coupled minimally to a scalar field with an arbitrary scalar potential  is taken as
\begin{equation} \label{action}
I[g, \varphi] = \frac{1}{16\pi G}\int d^{3}x\, \sqrt{-g} \Big(\CL_{EH}  +\CL_{\varphi}\Big)\,, 
\end{equation}
where
\[   
\CL_{EH}  = R - 2\Lambda\,, \qquad \CL_{\varphi} = - \frac{1}{2} \partial_{\mu}\varphi \partial^{\mu}\varphi  - V(\varphi)\,.
\]
In the following, we take $\Lambda =-1$ for our convenience and denote the relevant fields collectively as $\Psi$.

We consider the scalar hairy rotating black holes which have only one Killing vector. 
 Henceforth we adopt the metric and scalar fields ansatz as
\begin{align}
ds^2 &= -f(r,y)e^{2A(r,y)}dt^2 + \frac{dr^2}{f(r,y)}+r^2(d\theta -\O(r,y)dt)^2\,,\qquad y\equiv \Omega_{H} t -\theta\,, \\
\varphi & = \varphi(r,y)\,,
\end{align}
where $\Omega_{H}$ denotes the value of the metric function $\Omega$ at the outer horizon $r=r_{+}$. 
Generically, scalar hairy rotating $AdS_{3}$ black holes admit  only one Killing vector contrary to the Banados-Teitelboim-Zanelli(BTZ) case~\cite{Banados:1992wn}. 
Concretely, the above ansatz of metric and scalar fields admits a single Killing vector, which  is given by
\begin{equation} \label{}
\xi_{K} = \frac{\p}{\p t}  + \Omega_{H}\frac{\p}{\p \theta}\,.
\end{equation}
This becomes null at the horizon and thus  is the appropriate one for the entropy computation as a conserved charge.  Still, there are asymptotic Killing vectors $\xi_{T}\equiv \frac{\p}{\p t} $ and $\xi_{R}\equiv \frac{\p} {\p \theta}$, which can be used in order to define the total mass and angular momentum, respectively,  at the asymptotic infinity.

Let us compute the quasi-local Abbott-Deser-Tekin(ADT) charge~\cite{Abbott:1982jh,Deser:2002rt,Deser:2002jk,Kim:2013zha} for the above $\xi_{K}$ at the outer Killing horizon and check that it leads to the entropy as $\CS_{BH}$~\cite{Wald:1993nt,Iyer:1994ys,Wald:1999wa}.
It is very convenient to introduce the function $Z$ defined by
\begin{equation} \label{}
Z(r,y) \equiv r e^{A} f' + 2 r(e^{A})'f\,, \qquad  \quad {}' \equiv \frac{\p}{\p r}\,.
\end{equation}
Since the Noether potential for the Killing vector $\xi_{K}$  at the horizon is given by

\begin{equation} \label{}
K^{rt}(\xi_{K} ) = Z - r^{3}e^{-A}(\Omega - \Omega_{H})\Omega' \Big|_{r=r_{+}} = Z(r_{+})\,,
\end{equation}
one can see that~\cite{Wald:1993nt,Iyer:1994ys,Kim:2013zha,Kim:2013cor,Hyun:2014sha}
\begin{equation} \label{}
\frac{\kappa_H}{2\pi} \CS_{BH} = \frac{1}{16\pi G}\int_{\CH}dx_{\mu\nu}\, \Delta K^{\mu\nu} = \frac{1}{8G}Z(r_{+}) = \frac{1}{8G}r_{+}\,e^{A(r_{+})}f'(r_{+})\,,
\end{equation}
where $\kappa_H$ is the surface gravity at the horizon.
By recalling that the Hawking temperature and angular velocity at the horizon for the following ADM-decomposed metric 
\begin{equation} \label{}
ds^{2} =- N^{2}dt^{2}  + g_{ij}(dx^{i} + N^{i}dt)(dx^{j} + N^{j}dt)\,,
\end{equation}
are given by
\begin{equation} \label{}
T_{H} =\frac{\kappa_H}{2\pi} = \frac{1}{2\pi}\frac{\p_{r}N}{\sqrt{g_{rr}}}\bigg|_{r=r_{+}}\,, \qquad  \Omega_{H} = - N^{\theta}(r_{+})\,,
\end{equation}
 one can obtain the   temperature of these black holes as
\begin{equation} \label{}
  T_{H} = \frac{1}{4\pi r_{+}} Z(r_{+})= \frac{1}{4\pi}e^{A(r_{+})}f'(r_{+})\,. 
\end{equation}
In summary, we obtain the entropy of the hairy $AdS_{3}$ black holes in our setup as 
\begin{equation} \label{}
\CS_{BH} =  \frac{2\pi r_{+}}{4G}\,,
\end{equation}
which corresponds to  the standard form of the area law of black hole entropy. 

Generically,  the location of the outer and inner horizons are determined by the condition $f=0$. Near the outer horizon, the metric functions $f$, $e^A$ and $\Omega$ and the scalar field $\varphi$ may be expanded as
\begin{align}   \label{}
f(r,y)  &=  f'(r_{+})(r-r_{+})  + \cdots\,,  \qquad \qquad \qquad  e^{A(r,y)} = e^{A(r_{+})}\Big[1+A'(r_{+},y)(r-r_{+}) +\cdots\Big],  ~~~ \nn  \\  \\
 \Omega(r,y) &= \Omega_{H}+ \Omega'(r_{+},y) (r-r_{+}) +\cdots\,, \qquad \varphi (r,y)= \varphi_{H} (y)+ \varphi'(r_{+},y)(r-r_{+})+ \cdots\,. \nn
\end{align}
Note that the metric functions $f$, $e^{A}$ and $\Omega$ depend, in general, on the coordinates $r$ and $y$, but $f'(r_{+})$, $e^{A(r_{+})}$ and $\Omega_{H}$  are just constants, since the system under the consideration is assumed to have the bifurcating Killing horizon.  In other words, the metric function $f$ is independent of the coordinate $y$ up to the first order in the $(r-r_+)$ expansion.

At the asymptotic infinity, the asymptotic forms of metric functions can be expanded as
\begin{equation}  \label{asym} 
f (r, y) = r^{2}\Big[1+ \CO\Big(\frac{1}{r}\Big) \Big]\,, \qquad e^{A(r,y)} = 1 +\CO\Big(\frac{1}{r}\Big)\,, \qquad 
\Omega(r,y) = \frac{1}{r^{2}}\Big[\Omega_{\infty}(y) + \CO\Big(\frac{1}{r}\Big) \Big]\,. 
\end{equation}
By solving the scalar field equation, one can see that 
the scalar field takes  the  asymptotic behaviors as
\begin{equation} \label{scalarB}
\varphi(r,y) = \frac{\varphi_{-}(y)}{r^{\Delta_{-}}} +\cdots + \frac{\varphi_{+}(y)}{r^{\Delta_{+}}} + \cdots\,.
\end{equation}
Suppose that the fall off behavior of the scalar field is taken appropriately at the asymptotic infinity. Then it would be  sufficient to keep the scalar potential up to the quadratic term as $V(\f)=\frac{1}{2}m^{2}\f^2+\cdots$. This assumption gives us the scaling dimension as 
\begin{equation} \label{}
\Delta_{\pm} = 1 \pm \sqrt{1+m^{2}}\,.
\end{equation}
Note that the Breitenlohner-Freedman bound~\cite{Breitenlohner:1982jf} is given by $m^{2}=-1$ in our convention. 
Though $\varphi_{-}$ corresponds, usually,  to the non-normalizable mode, it becomes normalizable  for the range $-1< m^{2} < 0$   and, so both modes $\varphi_{\pm}$ may be turned on for the stable black hole solution~\cite{Hertog:2004dr,Hertog:2004bb}.    In this case, the constants $\Delta_{\pm}$  correspond to the dimensions of the dual operators. In the following, we restrict ourselves  on this range of the mass of the scalar field.   Our metric falloff boundary conditions  in Eq.~(\ref{asym}) correspond to the parameter range $-1<m^{2} \le -\frac{3}{4}$ and so we would like to focus on these cases in the following.  It would be possible to extend the range of the mass parameter $m^{2}$ by adjusting falloff boundary conditions appropriately. Furthermore, we focus on the case such as the difference between $\Delta_{-}$ and $\Delta_{+}$ is not integer and so that logarithmic modes do not appear in the solution. When the two modes are turned on generically, it has been known that the infinitesimal mass formula for the solution  is not  integrable.  This means that  the mass expression of hairy black hole solution may not be defined unambiguously. However, if there exists a relation between two modes $\varphi_{\pm}$, the finite mass expression can be obtained. Later on, we will obtain the mass of hairy black holes consistent with the scaling symmetry. In brief, it turns out that our mass expression corresponds to the one obtained by taking $\varphi_{+} =\nu \varphi_-^{\Delta_{+}/\Delta_{-}}$ with the dimensionless parameter $\nu$. 

\section{Scaling symmetry of the reduced action}
 In this section, we show the existence of the scaling symmetry in the reduced action for our ansatz and derive its conserved charge. 
The reduced action of Einstein gravity and  the scalar field parts in our ansatz becomes, respectively, 
\begin{align}
I_{red} &= \frac{1}{16\pi G} \int dr dy \,(L_{EH} + L_{\varphi})\,, \\
L_{EH} 
&= 2re^A  -e^A f' -\frac{e^A \dot{f}^2 }{2rf^2} - \frac{ \dot{(e^A)}\dot{f} }{rf}  + \frac{r^3\O'^2}{2e^A} + \frac{r\dot{f}\dot{\O}}{e^A f^2}(\Omega-\O_{H}) \,,  \\
L_{\varphi} 
&=-\frac{1}{2}r e^A \Big( 2V(\f)+ f \f'^2+\frac{ \dot{\f}^2}{ r^2} \Big) + \frac{r  \dot{\f}^2}{2 e^{A} f}(\O-\O_{H})^2 \,,
\end{align}
where ${\dot{}}$ denotes the derivative with respect to the coordinate $y$,~  $\dot{}\equiv\frac{\partial}{\partial y}$,  and $\Omega_{H}$ appears as the consequence of the derivative with respect to the coordinate $y$.  Note that the total derivative terms in the above reduced action are omitted, which are irrelevant in our computations. 
 Under a generic locally-supported variation, this reduced action transforms as
\begin{equation} \label{GenVar}
\delta I_{red} =  \frac{1}{16\pi G}\int dr dy\,  \Big[ \CE_{\Psi}\delta \Psi + \partial_a\Theta^a(\Psi\,;\,\delta \Psi) \Big]\,, \qquad a=r,y,
\end{equation}
where $\Psi$ denotes, collectively, various fields $f,A, \Omega$  and $\varphi$, and $\Theta^a$ are the total surface terms under the variation. It is straightforward to  obtain the equations of motion, $\CE_{\Psi}=0$,  by varying with respect  to  $f,\,A,\,\O$ and $\varphi$, which are given,  respectively, by
\begin{align}
0&=(e^A)' -\frac{r}{2}e^A\f'^2 -\frac{r\dot{\f}^2}{2e^Af^2}(\O-\O_{H})^2 + \frac{1}{rf}\Big(\dot{e^A}+ \frac{e^A \dot{f}}{f} \Big){\frac{\dot{}}{}}  -\frac{r}{f^2}\Big( e^{-A}\dot{\O}(\O-\O_{H}) \Big){\frac{\dot{}}{}} \,, \label{EOMf} \\[5pt]
0&= 2V(\f) + f\f'^2 +\frac{\dot{\f}^2}{r^2} + \frac{\dot{\f}^2}{e^{2A}f}(\O-\O_{H})^2  -4+\frac{2f'}{r} +\frac{3\dot{f}^2- 2f\ddot{f}}{r^2 f^2} +\frac{r^2\O'^2}{e^{2A}} \label{EOMA} \\
&\quad  +\frac{2\dot{f}\dot{\O}}{e^{2A}f^2}(\O-\O_{H}) \,,\nn\\[5pt]
0&=(r^{3}e^{-A}\Omega')' + \Big[\Big(\frac{r\dot{f}}{e^{A}f^{2}}\Big)\frac{\dot{}}{} + \frac{r\dot{\varphi}^{2}}{e^{A}f}\Big](\Omega-\Omega_{H})\,,
 \label{OmegaEOM}\\[5pt]
0& =(re^A f\f')' - re^A\frac{\partial V}{\partial \f} + \Big( \frac{e^A\dot{\f}}{r} -\frac{r\dot{\f}}{e^A f}(\O-\O_{H})^2 \Big){\frac{\dot{}}{}} \,. \label{EOMphi}
\end{align}
One can also see that the relevant surface term $\Th^a(\d f,\delta A\,,\d\O,\,\d\phi)$  is given by
\begin{align}
\Th^r &= -e^A \d f+ r^3 e^{-A}\O' \d\O - re^A f\varphi' \d\varphi \,, \\
\Theta^{y} &= \Big(\frac{r}{f^{2}}e^{-A}(\Omega-\Omega_{H}) \dot{\Omega}-\frac{(e^{A}f)}{rf^{2}}{\frac{\dot{}}{}} \Big)\delta f -\frac{e^{A}\dot{f}}{rf}   \delta A+ \frac{re^{-A}\dot{f}}{f^{2}}(\Omega -\Omega_{H}) \delta \Omega \nn \\
&\quad +\Big(\frac{re^{-A}(\Omega-\Omega_{H})^{2}}{f} - \frac{e^{A}}{r}\Big)\dot{\varphi}\delta \varphi \,. \nn
\end{align}

Now, let us consider the scaling symmetry of the above reduced action. This can be achieved by the appropriate weight assignment to various fields and their variations. 
The assignment of weights for various fields in the above reduced action is 
\begin{align}   \label{}
\delta_{\sigma}f&= \sigma(2f - rf')\,, \qquad  
\delta_{\sigma}e^{A} = \sigma\Big(-2e^{A} - r(e^{A})' \Big)\,,   \\
\delta_{\sigma}\Omega &=  \sigma(-2\Omega - r \Omega')\,,  \qquad 
\delta_{\sigma}\Omega_{H} = -2\s\Omega_{H}\,, \qquad 
\d_\s \f =-\s r\f' \,,\qquad\nn
\end{align}
and the derivative with respect to $y$ increases the weight by one, as
\begin{align} \label{}
\delta_{\sigma}\dot{f} &= \sigma(3\dot{f} - r\dot{f}')\,, \qquad 
\delta_{\sigma}\dot{e^A} = \sigma\Big(- \dot{e^A} - r (\dot{e^A})' \Big)\,,  \\
\delta_{\sigma}\dot{\Omega} &= \sigma(-\dot{\Omega} - r\dot{\Omega}')\,, \qquad \d_\s \dot{\f}=\s( \dot{\f}- r\dot{\f}') \,.\nn
\end{align}
Under this scaling transformation, the reduced Lagrangian transforms as
\begin{align}
\d_\s L_{\f} = \s( -L_{\f} -rL_{\f}' ) \,,\qquad
\d_\s L_{EH} = \s( -L_{EH} -rL_{EH}' )\,,
\end{align}
hence the reduced action is invariant up to total derivatives,
\begin{align} \label{SymVar}
\d_\s I_{red}=\frac{1}{16\pi G}\int dr dy \, \partial_a S^a\,, \quad S^r=-r(L_{\f}+L_{EH}), \quad S^y=0\,.
\end{align}

One may introduce the `would-be' Noether current for the scaling symmetry
\begin{align}
C^a &\equiv\Th^a - S^a\,.
\end{align}
Since we have assigned the coordinate $y$ to have the weight $-1$ and the parameter $\Omega_{H}$ to transform under the scaling symmetry, the conventional Noether procedure has  some subtleties, which need to be spelled out explicitly. Because of the unusual aspect of the scaling transformation $\delta_{\sigma}$ that it does not commute with the derivative with respect to $y$,  it  violates the locally-supported variation property in Eq.~(\ref{GenVar}). The other unusual aspect such that the parameter $\Omega_{H}$ transforms under the scaling symmetry  also violates such a property. Therefore, the scaling  transformation of the reduced action becomes 
\begin{align}
\delta_{\sigma} I_{red} =  \frac{1}{16\pi G}\int dr dy\,  \Big( \CE_{\Psi}\delta_\sigma \Psi
+ \partial_a\Theta^a(\Psi\,;\,\delta_\sigma \Psi)  + \frac{\delta I_{red}}{\delta \dot\Psi}\Big[\delta_{\sigma},\frac{\p}{\p y}\Big]\Psi + \frac{\delta I_{red}}{\delta \Omega_H}\delta_{\sigma}\Omega_H\Big) \,.
\end{align}
It is straightforward to see that $[\delta_{\sigma},\frac{\p}{\p y}]\Psi = \dot{\Psi}$ from the scaling  transformation and
\begin{align}   \label{Add}
\frac{\delta I_{red}}{\delta \dot\Psi} \dot{\Psi}&= 2 \bigg[ -\frac{e^A \dot{f}^2 }{2rf^2} - \frac{ \dot{(e^A)}\dot{f} }{rf} + \frac{r\dot{f}\dot{\O}}{e^A f^2}(\Omega-\Omega_{H}) - \frac{  e^A \dot{\f}^2}{2 r} + \frac{r  \dot{\f}^2}{2 e^{A} f}(\Omega-\Omega_{H})^2 \bigg] \,,    \\
\frac{\delta I_{red}}{\delta \Omega_H} &= -\frac{r }{e^{A}f^{2}}\Big[ f \dot{\varphi}^2 (\Omega -\Omega_{H}) + \dot{f}\dot{\O} \Big]\,. 
\end{align}
Identifying the above variation with the scaling variation given in Eq.~(\ref{SymVar}) and using the definition of the current $C^{a}$, one obtains 
\begin{equation} \label{extra}
\partial_{a}C^a = 
\Big[- \frac{\delta I_{red}}{\delta \dot\Psi} \dot{\Psi} + 2\Omega_{H}\frac{\delta I_{red}}{\delta \Omega_H}\Big]_{on-shell} \,,
\end{equation}
where we have used the on-shell condition $\CE_{\Psi}=0$.

 Let us define the  charge function for the scaling symmetry as
\begin{align}
C(r) =\frac{1}{16\pi G}\int^{2\pi}_{0} dy \,C^r \,,
\end{align}
which is not warranted to be conserved. Note that the $r$ component of the  current is given by
\begin{align}
C^r \label{Cr}
&=  -e^A(2f-rf') +r^2e^A f\f'^2 -r^3e^{-A}(\O^2)'  \nn \\
&~~~\,\, +\frac{r^2e^{-A} \dot{\f}^2}{ f}(\O-\O_{H})^2 +\frac{2r^2e^{-A}\dot{f}\dot{\O}}{f^2}(\O-\O_{H}) - \Big( \frac{e^A \dot{f}}{f} \Big)\frac{\dot{}}{}\,,
\end{align}
where we have used  Eq.~(\ref{EOMA}) to remove the scalar potential $V(\f)$ in the expression.
Though the current $C^{a}$ is not conserved even at the on-shell, one can apply the divergence theorem  to the above expression $\p_{a}C^{a}$ and obtains the following  result:
\begin{equation} \label{ChargeC}
C(r\rightarrow\infty) - C(r_{+})= \int drdy  \Big[- \frac{\delta I_{red}}{\delta \dot\Psi} \dot{\Psi}  + 2\Omega_{H}\frac{\delta I_{red}}{\delta \Omega_H} \Big]\,.
\end{equation}
%

\section{The Smarr relation}
In this section, we  compute the quasi-local ADT  charges at the horizon and at the asymptotic infinity for asymptotic Killing vectors and find the relation among these charges through the relation given in Eq.~(\ref{ChargeC}) for the  charge function $C(r)$. This relation gives us the novel Smarr relation, which reduces to the conventional one whenever  the scalar hair is turned off. Our results extend  those for the spherically symmetric case given in~\cite{Banados:2005hm,Hyun:2015tia}.

 At the horizon located at $r=r_{+}$, the value of the charge function $C(r)$ is given by
\begin{equation} \label{}
C(r_{+}) = \frac{1}{16\pi G}\int dy\Big[r_{+}e^{A(r_{+})}f'(r_{+}) -2r^{3}_{+}e^{-A(r_{+})}\Omega_H \Omega'(r_{+},y) \Big]\,.
\end{equation}
In order to consider the Smarr relation in this case, one need to use the near horizon angular momentum in matching with the value of the charge function at the horizon, $C(r_{+})$ from the scaling symmetry of the reduced action.  Explicitly, the near horizon geometry in our ansatz becomes 
\begin{align}   \label{}
ds^{2}_{NH} &= -e^{2A(r_{+})}\Big[1+A'(r_{+},y)(r-r_{+})\Big] f'(r_{+})(r-r_{+})dt^{2} + \frac{dr^{2}}{f'(r_{+})(r-r_{+})}  \nn \\
& + \Big(r^{2}_{+}+2r_{+}(r-r+)\Big)\Big[d\theta - \Big(\Omega_{H}+\Omega'(r_{+},y)(r-r_{+})\Big) dt\Big]^{2}\,,
\end{align}
on which, $\xi_{R}\equiv \frac{\p}{\p \theta}$ becomes an asymptotic Killing vector in the sense that
\begin{equation} \label{asymH}
\Lie_{\xi_R}g= 2 \nabla_{(\mu}\xi^{R}_{\nu)} =\CO\Big( (r-r_{+})\Big) \,.
\end{equation}
One can apply the formalism for asymptotic Killing vectors presented in Ref.~\cite{Hyun:2014kfa,Hyun:2014sha} to obtain the conserved charge $J_{H}$ for $\xi_{R}$ on the horizon. The upshot of the asymptotic Killing vector case is the existence of the additional term ${\bf A}$ in the quasi-local ADT potential as
\begin{equation} \label{}
2\sqrt{-g} {\bf Q}^{\mu\nu}_{ADT} = \delta K^{\mu\nu} -2\xi^{[\mu}\Theta^{\nu]} + \sqrt{-g} {\bf A}^{\mu\nu} (\Lie_{\xi} \Psi,\delta \Psi)\,,
\end{equation}
where the additional term ${\bf A}$ is composed of two parts ${\bf A}_{g} +{\bf A}_{\varphi}$, one of them, ${\bf A}_{g}$ is given by
\[   
{\bf A}^{\mu\nu}_{g}(\Lie_{\xi}g, \delta g) = -\Big(g^{\mu(\alpha}g^{\beta)(\rho}g^{\sigma)\nu} - g^{\nu(\alpha}g^{\beta)(\rho}g^{\sigma)\mu}\Big)\Big(\Lie_{\xi}g_{\alpha\beta}\delta g_{\rho\sigma} - \delta g_{\alpha\beta}\Lie_{\xi}g_{\rho\sigma}\Big)\,,
\]
as given in Ref.~\cite{Hyun:2014kfa} and the other one ${\bf A}_{\varphi}$ is given in the appendix B. However, the additional term does not contribute to the charge on the horizon since $\Lie_{\xi}g$ vanishes  asymptotically as was given in Eq.~(\ref{asymH}). The horizon angular momentum, which is the conserved charge for the asymptotic Killing vector $\xi_{R}$, can be computed as 
\begin{align}   \label{}
J_{H} &= -\frac{1}{8\pi G}\int ds \int_{\CH} dx_{\mu\nu}\sqrt{-g}{\bf Q}^{\mu\nu}_{ADT}(\xi_{R}\,;\, \delta_{s} \Psi )  \nn \\
&= -\frac{1}{16\pi G}\int_{\CH} dx_{\mu\nu}\Delta K^{\mu\nu}  \nn \\
&=-\frac{1}{16\pi G} \int dy~  r^3_{+} e^{-A(r_{+})}\O'(r_{+},y)\,,
\end{align}
where we have introduced the overall minus sign for the angular momentum. This convention is consistent with the one in Ref.~\cite{Iyer:1994ys}. 
Now, it is straightforward to check that 
\begin{equation} \label{SmarrH}
T_{H}\CS_{BH} +  2\Omega_{H}J_{H} = C(r_{+})\,,
\end{equation}
where $J_{H}$ is the angular momentum at the horizon.

Similarly, the total angular momentum may be computed at the asymptotic infinity through the asymptotic Killing vector $\xi_{R} \equiv \frac{\p}{\p\theta}$ and is given by 
\begin{align}   \label{}
J_{\infty} &= -\frac{1}{8\pi G}\int ds \int_{r\rightarrow\infty} dx_{\mu\nu}\sqrt{-g} {\bf Q}^{\mu\nu}_{ADT}(\xi_{R}\,;\, \delta\Psi)  \nn \\
&= -\frac{1}{16\pi G}\int_{r\rightarrow\infty} dx_{\mu\nu}\Big[\Delta K^{\mu\nu} + \int ds \sqrt{-g} {\bf A}^{\mu\nu} \Big] \nn \\
&=-\frac{1}{16\pi  G} \int dy~ r^3 e^{-A}\O'|_{r\rightarrow\infty}\,,
\end{align}
which does not need to be identical with the angular momentum at the horizon, $J_{H}$.

Some comments are in order.   The total angular momentum and the horizon angular momentum are identical, $J_{\infty} = J_{H}$ when the scalar hair is turned off.  For example, the angular momentum of BTZ black holes is given by $J^{BTZ} = r_{+}r_{-}/4G$ in our convention. Since the metric functions $\Omega$ and $A$ in the hairless BTZ black hole case are given by 
\[   
\Omega = \frac{r_{+}r_{-}}{r^{2}}\,, \qquad A =0\,,
\]
the value of the combination $r^{3}e^{-A}\Omega'$ in our computation becomes the same at the horizon and at the asymptotic infinity. In fact, one can show that the angular momentum $J$ may be defined at the arbitrary position of $r$, since $\xi_{R}$ is the exact Killing vector on the BTZ black hole geometry and the value of $r^{3}e^{-A}\Omega'$ is invariant along the radial direction. This is the special case of the general results on the angular momentum in the pure Einstein gravity~\cite{Julia:1998ys,Kim:2013zha}. However, it does not need to be the case with a scalar hair. 
By using the equations of motion given in Eq.~(\ref{OmegaEOM}) with the integration by parts, one can see that
\begin{equation} \label{DelJ}
\Delta J \equiv J_{H} - J_{\infty} = \frac{1}{16\pi G}\int d y \Big[r^{3}e^{-A}\Omega'\Big]^{r=\infty}_{r=r_{+}} = - \frac{1}{16\pi G}\int drdy \frac{\delta I_{red}}{\delta \Omega_H} \,.
\end{equation}
%
%
%
%
%
%

Let us consider the value of the charge function $C$ at the asymptotic infinity. By using the asymptotic behaviors of the metric functions and the scalar field given in Eq.~(\ref{asym}), one can see that the value of the charge $C$ at the asymptotic infinity  depends only on the first two terms in Eq.~(\ref{Cr}) as 
\begin{equation} \label{}
C(r\rightarrow\infty) = \frac{1}{16\pi G} \int dy \Big[ - e^A(2 f - r f')
+ r^2 e^{A} f \varphi'^2    \Big]\,.
\end{equation}
In order to relate this charge with the total conserved quantity, let us consider the total mass of scalar hairy $AdS_3$ black holes. 
The infinitesimal form of the mass of the hairy rotating black holes may be computed for the asymptotic Killing vector $\xi_{T} \equiv \frac{\p}{\p t}$,  just like the asymptotic angular momentum case, as
\begin{align} \label{delM}
\delta M_{\infty} &= \frac{1}{8\pi G} \int_{r\rightarrow\infty} dx_{\mu\nu}\sqrt{-g}\, {\bf Q}^{\mu\nu}_{ADT}(\xi_{T}\,;\, \delta \Psi\,)  \nn \\
&= \frac{1}{16\pi G}\int_{r\rightarrow\infty} dx_{\mu\nu} \Big[ \delta K^{\mu\nu}(\xi_{T})  - 2 \xi^{[\mu}_{T}\Theta^{\nu]} + \sqrt{-g}{\bf A}^{\m\n}\Big]  \nn \\
&= \frac{1}{16\pi G}\int dy \Big[ - e^A\d f - r e^A f \f' \d\f  -  r^3 \O \d(e^{-A}\O') \Big]_{r\rightarrow\infty}\,,
\end{align}
where the term $\sqrt{-g}{\bf A}(\Lie_{\xi_{T}}\Psi,\delta \Psi) $ turns out to vanish as $r\rightarrow\infty$. 
In the one-parameter path integration to obtain the finite mass expression of scalar hairy $AdS_3$ black holes,  the explicit form of solutions with a certain parameter, at least asymptotically, is required. 
As was done in Ref.~\cite{Hyun:2015tia}, it is very useful to obtain the finite mass expression in another way, which is quite appropriate  to compare it with the expression of the charge $C$. To this purpose, we introduce the `on-shell' scaling transformation under which the mass of hairy black holes transforms definitely with a certain weight. 

To simplify the presentation in the following, we assume that the  `on-shell' scaling transformation can be implemented on the solution. For instance,  this assumption is satisfied trivially, when one of $\varphi_{\pm}$ modes is turned off. When neither of them vanish, the assumption is valid but the meaning of the `on-shell' scaling transformation is related to the integrability issue on the scalar contribution, which is relegated to the next section.  
 In the present case, the `on-shell' scaling transformation can be implemented as
\begin{align}   \label{}
\hat{\delta}_{\sigma}f &=\sigma( 2f -rf') \,, \qquad \hat{\delta}_{\sigma}e^{A} = -\sigma r (e^{A})'\,, \nn \\
\hat{\delta}_{\sigma} \Omega &= -\s r\Omega'\,, \qquad \qquad \quad 
\hat{\delta}_{\sigma}\varphi = -\s r\varphi'\,. 
\end{align}
This `on-shell' scaling transformation corresponds to the specific choice of the one parameter path in the solution space, which becomes `integrable' in the sense of the parameter variation and so leads to the finite mass expression. Since this specific choice of parameter path might not be realized in the most generic falloff boundary conditions allowed in the range $ -1 < m^{2} \le -\frac{3}{4}$, we restrict ourselves to  the class of solutions  in which such a path is allowed.  Nevertheless, this encompasses most of the interesting cases explored in the literatures on three-dimensional hairy black holes~\cite{Hyun:2015tia}.

Since  the  black holes under consideration are asymptotically AdS, the `on-shell' scaling leads to $\hat{\delta}_{\sigma}M_{\infty} = 2\sigma M_{\infty}$~\cite{Hyun:2015tia}.
By using  the following  `on-shell' scaling transformation of the function,
$ e^{-A}\Omega'$,
\[   
\hat{\delta}_{\sigma} (e^{-A}\Omega')= \sigma \Big[ -e^{-A}\Omega' -r(e^{-A}\Omega')' \Big]\,,
\]
and the asymptotic behaviors of the metric functions $e^{A}$ and $\Omega$ given in Eq.~(\ref{asym}), one can see that
\begin{equation} \label{}
\hat{\delta}_{\sigma} (e^{-A}\Omega') \Big|_{r\rightarrow\infty} =\CO\Big(\frac{1}{r^{3}}\Big)\,.
\end{equation}
Then, the mass of hairy black holes can be obtained from this `on-shell' scaling transformation as
\begin{equation} \label{finiteM}
2\sigma M_{\infty} =  \hat{\delta}_{\sigma} M_{\infty}  = \frac{\s}{16\pi G}\int dy \Big[ - e^A(2f-rf') + r^{2}e^A f \varphi'^{2}\Big]_{r\rightarrow\infty}\,.
\end{equation}
where $\hat{\delta}_{\sigma} M$ is computed from Eq.~(\ref{delM}).
%
%
As a result, it is straightforward to check that 
\begin{equation} \label{}
2M_{\infty}  = C(r\rightarrow\infty)\,.
\end{equation}
Note that we have assumed that no additional dimensionful parameter appears in the solution. Whenever another dimenionful parameter exists, we need to compensate its variation under the `on-shell' scaling, as was done in Ref.~\cite{Hyun:2015tia}. More details are given in the next section.

Combining the above result with the relation among the conserved charge at the horizon given in Eq.~(\ref{SmarrH}) and the relation given in Eq.~(\ref{ChargeC}), one can obtain the following novel Smarr relation  
for scalar hairy $AdS_{3}$ black holes 
\begin{align}   \label{}
C(r\rightarrow\infty) - C(r_{+})&=  2M_{\infty} - T_{H}\CS_{BH} - 2\Omega_{H}J_{H}  \nn \\
&=  -\frac{1}{16\pi G}\int dr dy  \Big[\frac{\delta I_{red}}{\delta \dot\Psi}\dot{\Psi}   - 2\Omega_{H}\frac{\delta I_{red}}{\delta \Omega_H} \Big]_{on-shell}\,.
\end{align}
Note that the angular momentum in the conventional  Smarr relation is taken as the one, $J_{\infty}$ defined at the asymptotic infinity. 
By noting the relation given in Eq.~(\ref{DelJ}), one can see that our Smarr relation can also be written as
\begin{equation} \label{}
  2M_{\infty} - T_{H}\CS_{BH} - 2\Omega_{H}J_{\infty}  =  -\frac{1}{16\pi G}\int dr dy  \Big[\frac{\delta I_{red}}{\delta \dot\Psi}\dot{\Psi}\Big]_{on-shell} \,.
\end{equation}

We would like to mention that the charge function $C(r)$ depends on the choice of coordinates and so the identification of its value with  conserved charges may depend on the choice of coordinates. 
As is shown explicitly in our example of scalar hairy $AdS_{3}$ black holes  with  the choice of coordinates  such as $\Omega(r_{+}) = \Omega_{H}$,  the angular momentum, $J_{H}$ defined on the near-hoizon space appears naturally  when  $C(r_{+})$ is identified with conserved charges of black holes.  In contrast, the charge function for  the coordinates such as $\Omega(r_{+})=0$  can be shown to lead to  the total angular momentum defined at the asymptotic infinity  when $C(r\rightarrow\infty)$ is identified with conserved charges of black holes. However, the final Smarr relation takes the same form if the conserved charges of black holes are consistently taken in these coordinates.
Our claim is  that  the relation for the charge function $C$ in Eq.~(\ref{ChargeC}) leads to the Smarr relation which should be coordinate independent. 


\section{The scalar hair preserving the AdS structure}
 In the previous section, we have adopted the finite mass expression of scalar hairy black holes at the asymptotic infinity, defined by the `on-shell' scaling path, to match it with the value of the charge function at the asymptotic infinity, $C(r\rightarrow\infty)$.  We would like to check that the finite mass expression for hairy black holes is consistent with the conventional one obtained by the integration along the  one parameter path in the solution space, which is not warranted to give the finite mass expression when the infinitesimal one is not integrable~\cite{Hertog:2004dr,Hertog:2004bb,Lu:2013ura,Lu:2014maa,Wen:2015uma}. Because of the non-integrability due to generic scalar boundary values, we need to impose a relation $\varphi_{+} = H(\varphi_{-})$ to obtain a finite mass expression~\cite{Hertog:2004dr,Hertog:2004bb,Hertog:2004ns}.

By solving the equations of motion asymptotically for the scalar field boundary condition given in Eq.~(\ref{scalarB}), one obtains 
the  metric functions in the form of 
\begin{align}
f(r,y)&=r^{2}\bigg[ 1 - \frac{1}{r^{2}}\Big(b_0 + c_{1}(y) \Big)  + \frac{\Delta_{-}\varphi_{-}^2(y)}{ 2\, r^{2\Delta_{-}}} +  \frac{\Delta_{+}\varphi^{2}_{+}(y)}{2\, r^{2\Delta_{+}}}+\cdots  \bigg]\,, \\[5pt]
e^A&= 1+ \frac{m^{2} \varphi_{-}(y) \varphi_{+}(y)}{2r^2}   - \frac{\Delta_{-}\varphi_{-}^2(y)}{ 4\, r^{2\Delta_{-}}} - \frac{\Delta_{+}\varphi^{2}_{+}(y)}{4\, r^{2\Delta_{+}}} +\cdots  \,, \\ 
\O&=\frac{1}{r^{2}}\Big(\o_0+c_2(y) \Big)+\CO\Big(\frac{1}{r^4}\Big) \,,
\end{align}
where $c_{1}(y)$ and $c_{2}(y)$ correspond to the back reaction of the metric functions to the scalar contribution and $b_{0}$ and $\omega_{0}$ do to the integration constants for the source-free equations of motion, up to the order expanded.  Note that we have denoted $  \omega_{0} + c_{2}(y)=\Omega_{\infty}(y)$ in Eq.~(\ref{asym}).  In the following, we do not need explicit expressions of the functions $c_{1,\,2}(y)$ to verify our claims.
%
%
%

By regarding $\varphi_{\pm}(y)$ and $c_{1,\,2}(y)$ as free parameters in the quasi-local ADT expression for the mass given in Eq.~(\ref{delM}), one can compute the quasi-local ADT mass at the asymptotic infinity for the asymptotic time-like  Killing vector $\xi_T=\frac{\partial}{\partial t}$ as
\begin{eqnarray}  \label{genericV}
\delta M_{\infty} &=&\frac{1}{16\pi G} \int dy \Big[\delta\Big(b_0+c_{1}(y) + \frac{1}{2}(\Delta_{+}+\Delta_{-})\varphi_{-}(y)\varphi_{+} (y) \Big)  \nn\\
&& \qquad \qquad\,\,\,\,\,\,\,\,+  \frac{1}{2}( \Delta_{+} - \Delta_{-})\Big (\varphi_{+}(y)  \d\varphi_{-} (y) -\varphi_{-}(y)\d\varphi_{+}(y) \Big)  \Big]\,. \end{eqnarray}
The second term clearly shows the non-integrability of $\delta M_{\infty}$ if $\varphi_\pm$ are independent each other. 
For any arbitrary given relation in the form of $\varphi_{+} =H(\varphi_{-})$,  we obtain
\begin{equation} \label{}
\delta M_{\infty} = \frac{1}{16\pi G} \int dy~ \delta \Big( b_0+ c_{1}(y)  + \Delta_{-}\varphi_{-}(y)\varphi_{+}(y) + (\Delta_{+}-\Delta_{-})\int^{\varphi_{-}(y)}_{0}H(\tilde{\varphi})d\tilde{\varphi}\Big)\,,
\end{equation}
which is a straightforward generalization of the result  in Ref.~\cite{Hertog:2004ns} for  the static case  with an exact Killing vector.

Now let us consider the boundary condition,  $H(\varphi_{-})=\nu \varphi^{\Delta_{+}/\Delta_{-}}_{-}$, which preserves the asymptotic AdS structure. Note that the parameter $\nu$ is dimensionless. It turns out  that the finite mass expression from the `on-shell' scaling transformation leads to  the same one from this boundary condition. 
By using this relation, we obtain
the finite mass expression  as 
\begin{equation} \label{}
M_{\infty} = \frac{1}{16\pi G} \int dy\Big[ b_0+ c_{1}(y)  -m^2\varphi_{-}(y)\varphi_{+}(y)\Big]\,.
\end{equation}

Let us confirm that the `on-shell' scaling transformation applied to the quasi-local ADT expression leads to the same result as the one from the above boundary condition. Whenever the scalar boundary condition is given only by one of $\varphi_{\pm}$, not both of them, it is straightforward to check that the `on-shell' scaling gives the same result.  When neither of $\varphi_{\pm}$ vanish, the complication arises because of the additional parameter $\nu$.  As was explained in~\cite{Hyun:2015tia},  the existence of a dimensionful  parameter requires us the compensating transformation to restore its invariance along the one parameter path in the solution space. However, the additional parameter, $\nu$, in this case is dimensionless, and so we do not need to consider the compensating transformation. Indeed, by inserting the above perturbative expansions of the metric functions and the scalar field  to the Eq.~(\ref{finiteM}), one obtains
 \begin{equation} \label{}
 \hat{\delta}_{\sigma}M_{\infty} = 2\sigma M_{\infty}= \frac{\sigma}{8\pi G} \int dy \Big[ b_0+ c_{1} (y)-m^2\varphi_{-}(y)\varphi_{+}(y)\Big]\,,
 \end{equation}
 which gives the same result  as the one with the above direct parameter variation method.  Therefore the Smarr relation and the first law follow in  the  forms as
\begin{align}    \label{SM1st}
 M_{\infty} &= \frac{1}{2}T_{H}\CS_{H} + \Omega_{H} J_{\infty}-\frac{1}{32\pi G}\int dr dy  \Big[\frac{\delta I_{red}}{\delta \dot\Psi} \dot{\Psi} \Big]\,,   \\
 dM_{\infty} &=  T_{H}d\CS_{H} + \Omega_{H}d J_{\infty}  \nn\,.
\end{align}
 %
 
\section{Thermodynamic stability}
In this section we provide a simple criterion for the thermodynamic instability of scalar hairy $AdS_{3}$ black holes which belong to the same branch with the BTZ black holes, which may not be applied to   some of scalar hairy black holes shown to belong to a different branch from the BTZ black holes~\cite{Correa:2010hf}.  The explicit example in this branch is studied recently in Ref.~\cite{Iizuka:2015vsa} by using a numerical technique, which we would like to understand from our novel Smarr relation.

To compare the thermodynamic stability between two black hole solutions in the conventional (grand) canonical ensemble, one need to place those solutions at the same temperature and at the same chemical potentials.  In our case, scalar hairy black holes have four free parameters,  which cannot be specified uniquely by fixing the temperature and the angular velocity of black holes.  Since it is daunting task to scan and compare all the possible configurations of hairy black holes with BTZ black holes  just by fixing two parameters through the choice of the temperature and the angular velocity, we would like to choose a specific configuration of hairy black holes by fixing additional free parameters coming from the scalar field. To make the choice, we would like to take  conserved charges of black holes as a fixing tool for the additional free parameters.  Succinctly speaking,
 the temperature and the angular velocity of hairy $AdS_{3}$  black holes and non-hairy BTZ ones are taken as the same value, and  in addition the values of mass and angular momentum of hairy black holes are also taken as the same values of BTZ ones as
\begin{equation} \label{SameV}
M^{hairy}_{\infty} = M^{BTZ}_{\infty}\,, \qquad T^{hairy}_{H} = T^{BTZ}_{H}\,, \qquad J^{hairy}_{\infty} =  J^{BTZ}_{\infty}\,, \qquad \Omega^{hairy}_{H}  = \Omega^{BTZ}_{H}\,.
\end{equation}
In canonical ensemble, one can compare the free energy of each solution to see which one is more stable in this setup. Because of our fixing conditions, the comparison in free energy is identical with the one in the entropy and so we focus on the entropy in the following.

Without resorting to the explicit solution, one can see the behavior of the entropy of each black hole by using the derived Smarr relation as follows. First, note that the derived Smarr relation may be rewritten in terms of the angular momentum defined at the asymptotic infinity as 
\begin{equation} \label{}
M_{\infty} = \frac{1}{2}T_{H}\CS_{BH} + \Omega_{H}J_{\infty} -\frac{1}{32\pi G}\int dr dy  \Big[\frac{\delta I_{red}}{\delta \dot\Psi} \dot{\Psi} \Big]\,.
\end{equation}
For hairy $AdS_3$ black hole solutions and non-hairy BTZ solutions, the Smarr relation takes the following form, respectively, as 
\begin{align}   \label{}
M^{hairy}_{\infty} &= \frac{1}{2}T^{hairy}_{H}\CS^{hairy}_{BH} + \Omega^{hairy}_{H}J^{hairy}_{\infty} -\frac{1}{32\pi G}\int dr dy  \Big[\frac{\delta I_{red}}{\delta \dot\Psi}  \dot{\Psi}\Big]\,,  \\
 M^{BTZ}_{\infty} &= \frac{1}{2}T^{BTZ}_{H}\CS^{BTZ}_{BH} +  \Omega^{BTZ}_{H}J^{BTZ}_{\infty}\,,
\end{align}
where the fact that $\frac{\delta I_{red}}{\delta \dot\Psi}  =0$ for BTZ black holes is used. 
%
Under the condition given in Eq.~(\ref{SameV}), one can see that
\begin{equation} \label{}
\CS^{hairy}_{BH} - \CS^{BTZ}_{BH}  = \frac{1}{T_{H}} \frac{1}{16\pi G}\int dr dy  \Big[\frac{\delta I_{red}}{\delta \dot\Psi}  \dot{\Psi}\Big]\,. 
\end{equation}
As a result, the relative thermodynamic  stability between hairy $AdS_{3}$ black holes and non-hairy BTZ ones is determined  by the sign of  the integral of  $\frac{\delta I_{red}}{\delta \dot\Psi} \dot{\Psi}$, which can be determined by inserting the explicit solution  to  the expression of $\frac{\delta I_{red}}{\delta \dot\Psi}\dot{\Psi}$ given in Eq.~(\ref{Add}).  It seems very plausible that  the integral is negative for the scalar potential consistent with the positive energy theorem since the no-hair theorem is argued to hold in such a case~\cite{Hertog:2006rr}.


\section{Another boundary condition of the scalar hair}
 %
 
In this section, we consider the case when a dimensionful parameter appears in the boundary condition of the scalar hair. The relation between $\varphi_{-}$ and $\varphi_{+}$ is taken  as $\varphi_{+} =H(\varphi_{-}) = \kappa \varphi_{-}$, in which the parameter $\kappa$ becomes dimensionful. This boundary condition corresponds to the double trace operator deformation in the dual conformal field theory.
In this case,  the generic result in Eq.~(\ref{genericV}) leads to the following infinitesimal mass expression 
\begin{equation} \label{kapV}
\delta M_{\infty}   = \frac{1}{16 \pi G}\int dy~  \delta \Big( b_0+ c_{1}(y)  +\varphi_{-}(y)\varphi_{+}(y) \Big)\,.
\end{equation}
It would be useful  to obtain this expression from the `on-shell' scaling transformation with the compensating term.
The `on-shell' scaling transformation leads to an unwanted transformation of the parameter $\kappa$  as 
\begin{equation} \label{}
\hat{\delta}_{\sigma}\kappa = \sigma( \Delta_{+} - \Delta_{-}) \kappa\,. 
\end{equation}
However, the parameter $\kappa$ should remain constant when we integrate along a one parameter path in the solution space. In other words, we should take the compensating transformation in the metric functions and the scalar field, up to the diffeomorphism transformation,  in the form, 
\begin{equation} \label{}
\hat{\delta}_{\kappa}\kappa = -\sigma( \Delta_{+} - \Delta_{-}) \kappa\,, \qquad 
\hat{\delta}_{\kappa}(\varphi_{+}\varphi_{-})=0
\end{equation}
By using the infinitesimal expression for the quasi-local ADT mass in Eq.~(\ref{delM}) or Eq.~(\ref{genericV}), one can see that
\begin{align}   \label{}
\hat{\delta}_{\kappa}M_{\infty} &= \frac{1}{16\pi G}\int dy~   \Big[    \frac{1}{2}( \Delta_{+} - \Delta_{-}) (\varphi_{+}  \hat{\delta}_{\kappa}\varphi_{-}  -\varphi_{-}\hat{\delta}_{\kappa}\varphi_{+} )  \Big]    \nn \\
 &= \frac{\sigma }{16\pi G}\int dy~  \Big[\frac{1}{2}(\Delta_{+}-\Delta_{-})^{2} \varphi_{-}(y)\varphi_{+}(y)\Big]\,.
\end{align}
%
%
%
In the end, when both of $\varphi_{\pm}$ are turned on,  the mass of scalar hairy black holes is given by
\begin{equation} \label{}
 \hat{\delta}_{\sigma}M_{\infty}+ \hat{\delta}_{\kappa}M_{\infty} = 2\sigma M_{\infty}= \frac{1}{8\pi G}\int dy~ \sigma\Big[ b_0+ c_{1}(y)  +\varphi_{-}(y)\varphi_{+}(y) \Big]\,,
\end{equation}
which is identical from the direct parameter variation, indeed.  We would like to emphasize that this derivation by the  `on-shell' scaling transformation  is intended to show that  it gives us the finite mass expression directly  and  is consistent with a certain boundary condition of scalar hair.

In the end, the finite mass expression in this boundary condition leads to the relation
\begin{equation} \label{}
C (r\rightarrow\infty)= 2(M_{\infty} - \mu_{\varphi} Q_{\varphi})\,, 
\end{equation}
where the scalar charge $Q_{\varphi}$ and its chemical potential $\mu_{\phi}$ are defined by 
\begin{equation} \label{}
\qquad  Q_{\varphi} \equiv \frac{1}{8\pi G}\int dy \Big[\varphi_{-}(y)\varphi_{+}(y) \Big]=\frac{\kappa}{8\pi G}\int dy \varphi_{-}^2(y)\,, \quad \mu_{\varphi} \equiv1+m^2\,.
\end{equation}
In this case, the Smarr relation and the first law are given by  
\begin{align}   \label{}
M_{\infty} &= \frac{1}{2} T_{H}\CS_{BH}  +\Omega_{H}J_{\infty} + \mu_{\varphi}Q_{\varphi} -\frac{1}{32\pi G}\int dr dy  \Big[\frac{\delta I_{red}}{\delta \dot\Psi}\dot{\Psi}\Big]_{on-shell} \,,   \\   dM_{\infty} &=  T_{H}d\CS_{BH}  +\Omega_{H}dJ_{\infty} \,. \nn
\end{align}
%
%
Just like the boundary condition considered in previous sections, one may propose the criterion for the thermodynamic stability of scalar hairy black holes as follows. At the same temperature and the angular velocity with the same total mass and angular momentum as in Eq.~(\ref{SameV}), one can see that
\begin{equation}
\CS^{hairy}_{BH} -\CS^{BTZ}_{BH} = \frac{2}{T_H}\bigg(\frac{1}{16\pi G}\int dr dy  \Big[\frac{\delta I_{red}}{\delta \dot\Psi}\dot{\Psi}\Big]_{on-shell}  - \mu_\varphi Q_{\varphi} \bigg)\,. 
\end{equation}
The thermodynamic stability of scalar hairy black holes relative to BTZ black holes  is determined by  the sign of the right hand side of the above equation. 
It has been argued~\cite{Faulkner:2010fh} that $\kappa$ should be negative for the existence of stable scalar hairy black holes, in which  $-\mu_{\varphi}Q_{\varphi} > 0$. Since the first term of the right hand side seems negative,  the stability of the hairy black holes is determined by the competition of those two terms.  The case studied in~\cite{Iizuka:2015vsa} seems to correspond to such a case that the second term $-\mu_{\varphi}Q_{\varphi}$ is dominant. It would be very interesting to confirm our claims through the numerical approach.

\section{Conclusion}

We would like to summarize what we have found and indicate some future directions. 
We have shown explicitly the novel Smarr relation for scalar hairy $AdS_{3}$ black holes under  the AdS-invariant boundary conditions is given by
\[   
M_{\infty} = \frac{1}{2}T_{H}\CS_{BH} +  \Omega_{H}J_{\infty} -\frac{1}{32\pi G}\int dr dy  \Big[\frac{\delta I_{red}}{\delta \dot\Psi}\dot{\Psi}\Big]_{on-shell} \,,
\]
which can be understood as a simple consequence of the existence of the scaling symmetry on the reduced action. On the other hand, when the boundary conditions preserve partially AdS structures,  we have shown that the dimensionful parameter $\kappa$ appears and the Smarr relation contains the scalar charge $Q_{\varphi}$ as
\[   
M_{\infty} = \frac{1}{2} T_{H}\CS_{BH}  +\Omega_{H}J_{\infty} + \mu_{\varphi}Q_{\varphi} -\frac{1}{32\pi G}\int dr dy  \Big[\frac{\delta I_{red}}{\delta \dot\Psi}\dot{\Psi}\Big]_{on-shell} \,.
\]

This Smarr relation is derived without an explicit analytic solution and holds generically for all kinds of black holes satisfying our ansatz.  
In contrast to the Smarr relation,  the first law of black hole thermodynamics holds for both boundary conditions,  in the form of 
\[   
dM_{\infty} = T_{H}d\CS_{BH} + \Omega_{H} dJ_{\infty}\,,
\]
which is a consequence of the Stokes' theorem.  

Since there are integrability issues when black hole solutions possess two independent parameters  $\varphi_{\pm}$ originated from the two independent scalar modes, we have explicitly verified our total mass expression $M_{\infty}$  by using `on-shell' scaling method and  by taking a specific boundary condition, which preserves asymptotic AdS structure, in the parameter variation method. We have also derived the finite mass expression when the dimesnsionful parameter appears in the boundary conditions of the scalar field.

We have also shown that the thermodynamic  stability of scalar hairy black holes, compared with the one without the deformation due to scalar hairs, can be determined by the sign of the integral value of $\frac{\delta I_{red}}{\delta \dot{\Psi}}\dot{\Psi}$ for the AdS-invariant boundary conditions. In contrast, it is determined by the competition between the integral value and the scalar charge contribution $-\mu_{\varphi}Q_{\varphi}$ for the other boundary conditions. This gives us a very simple criterion for the existence of stable scalar hairy black holes from the thermodynamic viewpoint. 
 It would be very interesting to check our claims by using numerical methods    
 and  to extend our results in the higher dimensional case.

\vskip 1cm
\centerline{\large \bf Acknowledgments}
\vskip0.5cm
{SH was supported by the National Research Foundation of Korea(NRF) grant 
with the grant number NRF-2013R1A1A2011548. SY was supported by the National Research Foundation of Korea(NRF) grant 
with the grant number NRF-2015R1D1A1A09057057. }

{\center \section*{Appendix}}

\section*{A. Equations of motion}
\renewcommand{\theequation}{A.\arabic{equation}}
  \setcounter{equation}{0}

%
We display relevant original scalar and metric equations of motion in this appendix. The Euler-Lagrange expressions for the scalar and metric fields are defined by
\begin{align}
\CE_\f &\equiv \Box\f - \frac{\partial V(\f)}{\partial \f}\,, \\
  \CE_{\mu\nu} &\equiv G_{\mu\nu} + \Lambda g_{\mu\nu} - T^\varphi_{\mu\nu}\,, \qquad T^\varphi_{\mu\nu} \equiv \frac{1}{2}\p_\mu\varphi \p_\nu \varphi - \frac{1}{4}g_{\mu\nu}\Big(\p_\alpha \varphi \p^{\alpha}\varphi + 2V(\varphi) \Big)\,,
\end{align}
where $G_{\mu\nu}$ is the Einstein tensor.
The Euler-Lagrange expression of  the scalar field in our ansatz becomes
\begin{align}
\CE_\f 
&= A' f \varphi' + f' \varphi' +\frac{f \varphi'}{r} + f \varphi''
- \frac{\partial V(\f)}{\partial\f} 
+\frac{\dot{A} \dot{\varphi} + \ddot{\varphi}}{r^2} \nn \\[5pt]
&\quad+ \frac{(\O-\O_H)^2 }{e^{2A} f }\bigg[ \dot{A}\dot{\f} + \frac{\dot{f}\dot{\f}}{f} -\ddot{\f} \bigg] - \frac{2\dot{\f}\dot{\O}(\O-\O_H) }{e^{2A}f} \,.
\end{align}
Three independent metric Euler-Lagrange expressions among $\CE_{\mu\nu}$ in our ansatz are
\begin{align}
\CE^{t}\,_{t}-\CE^{r}\,_{r} &= \frac{1}{2} f \varphi'^2 - \frac{fA'}{r} - \frac{r^2\O}{2 e^{2 A}} \bigg[ A' \Omega' -  \Omega'' -  \frac{3\Omega'}{r}  \bigg] -\frac{\dot{A}^2 + \ddot{A}}{r^2} + \frac{1}{r^2 f}\bigg[ -  \dot{A} \dot{f} -\ddot{f} + \frac{\dot{f}^2}{f} \bigg] \nn\\[5pt]
&\quad
-\frac{ \Omega(\O-\O_H) }{2 e^{2 A}f^2} \bigg[ \dot{A} \dot{f} + \frac{\dot{f}^2}{f} - \ddot{f} + \frac{\O_H}{\O}f\dot{\varphi}^2 \bigg]
-\frac{(\Omega-\O_H) }{e^{2 A}f} \bigg[ \dot{A} \dot{\Omega}-\ddot{\Omega} \bigg] +\frac{ \dot{\Omega}^2}{e^{2 A}f}\,,
\\ \nn\\
\CE^{t}\,_{\th}+ \O\,\CE^{\th}\,_{\th} &= \Omega \bigg[ -1 + \frac{V}{2} +\frac{ f'}{2r} + \frac{f \varphi'^2}{4}  \bigg]
-\frac{f}{2} \bigg[  A' \Omega' - \Omega'' - \frac{3\Omega'}{r} \bigg] \nn\\[5pt]
&\quad
-\frac{1}{4r^2 f} \bigg[ 2  \dot{A} \dot{f}(\Omega-\O_H) +\frac{\dot{f}^2 }{f}(\O-4\O_H) +  2\O_H  \ddot{f} +  f\dot{\varphi}^2 (\O-2\O_H) \bigg]
\nn\\[5pt]
&\quad
+ \frac{ \O }{4e^{2 A}} \bigg[ r^2\O'^2 +\frac{2\dot{f} \dot{\Omega}}{f^2} (\O-\O_H) + \frac{\dot{\varphi}^2}{f}(\O-\O_H)^2 \bigg] \,,
\\ \nn \\
\CE^{\th}\,_{t} &= \frac{r^2}{2e^{2 A}}  \bigg[ A' \Omega' - \Omega'' - \frac{ 3\Omega'}{r} \bigg]
+\frac{ (\Omega-\O_H) }{2 e^{2 A} f^2} \bigg[\dot{A} \dot{f} +\frac{\dot{f}^2}{f} -\ddot{f} + f\dot{\varphi}^2 \bigg] \,.
\end{align}
One can check that all these equations of motion, $\CE_\f=0$ and $\CE^\mu\,_\nu=0$, are completely consistent with Eqs.~(\ref{EOMf}) $\sim$ (\ref{EOMphi}).

\subsection*{B.  Some useful formulae}
\renewcommand{\theequation}{B.\arabic{equation}}
  \setcounter{equation}{0}
%
In this appendix, we derive the additional contribution of the scalar field to the quasi-local ADT charges for  an asymptotic Killing vector. 
To preserve the  off-shell conservation property of the current for the asymptotic Killing vector $\zeta$, one need to add an additional term $\CJ^{\mu}_{\Delta}$ to the current  $\CJ^{\mu}_{ADT}$ for the exact Killing vector, as
\begin{equation} \label{}
{\bf J}^{\mu}_{ADT} = \CJ^{\mu}_{ADT} + \CJ^{\mu}_{\Delta}\,,
\end{equation}
where the ADT current is defined by (see~\cite{Hyun:2014kfa,Hyun:2014sha} for some details.)
\[   
\sqrt{-g}\CJ^{\mu}_{ADT} = \delta (\sqrt{-g}{\bf E}^{\mu\nu}\zeta_{\nu}) + \frac{1}{2}\sqrt{-g}\zeta^{\mu}\CE_{\Psi}\delta \Psi\,.
\]
Since the metric part of the additional current $\CJ^{\mu}_{\Delta}$ was already considered in~\cite{Hyun:2014kfa}, let us focus on the scalar part of $\CJ^{\mu}_{\Delta}$, which is denoted  as $\CJ^{\mu}_{\Delta,\, \varphi}$. Repeating the computation in~\cite{Hyun:2014kfa} with a scalar field, it is a straightforward exercise to obtain the following expression 
\begin{equation} \label{}
2\p_{\mu}(\sqrt{-g}\CJ^{\mu}_{\Delta, \varphi}) = \delta(\sqrt{-g}\CE_{\varphi})\Lie_{\zeta}\varphi  -  \Lie_{\zeta}(\sqrt{-g}\CE_{\varphi})\delta\varphi  +\Lie_{\zeta}(\sqrt{-g}T^{\varphi}_{\mu\nu})\delta g^{\mu\nu} - \delta(\sqrt{-g}T^{\varphi}_{\mu\nu})\Lie_{\zeta}g^{\mu\nu} \,.
\end{equation}
By applying the integration by parts iteratively on the following quantity 
\begin{equation} \label{}
\delta (\sqrt{-g}\CE_{\varphi}) \Lie_{\zeta}\varphi- \delta(\sqrt{-g}T^{\varphi}_{\mu\nu})\Lie_{\zeta}g^{\mu\nu}\,, 
\end{equation}
one can see that
\begin{align}   \label{}
2\CJ^{\mu}_{\Delta,\varphi} &= \frac{1}{2}(g^{\mu\alpha}g^{\nu\beta} +g^{\nu\alpha}g^{\mu\beta} - g^{\alpha\beta}g^{\mu\nu} )\nabla_{\nu}\varphi ( \Lie_{\zeta}g_{\alpha\beta}\delta \varphi - \delta g_{\alpha\beta}\Lie_{\zeta} \varphi) 
+   \Lie_{\zeta}\varphi \nabla^{\mu}\delta\varphi - \delta\varphi\nabla^{\mu}\Lie_{\zeta}\varphi \,.
\end{align}

Just as in the ADT current, the ADT potential for the asymptotic Killing vector has an additional anti-symmetric tensor term ${\bf A}^{\mu\nu} = {\bf A}^{\mu\nu}_{g} + {\bf A}^{\mu\nu}_{\varphi}$ as
\begin{equation} \label{}
2\sqrt{-g}{\bf Q}^{\mu\nu}_{ADT} (\zeta, \Psi) = \delta K^{\mu\nu} - 2\zeta^{[\mu}\Theta^{\nu]} + \sqrt{-g}{\bf A}^{\mu\nu}\,.
\end{equation}
Though there is an additional  term to the ADT current, the additional term to the ADT potential, ${\bf Q}^{\mu\nu}_{ADT}$ from the scalar field can be shown to vanish in our case. This result may be derived by following the same procedure for the metric part given in~\cite{Hyun:2014kfa}.
From the surface term of the action, given in Eq.~(\ref{action}), for the scalar field $\varphi$ variation 
\begin{equation} \label{}
\delta\Theta^{\mu}(\Lie_{\zeta}\varphi) = -\delta\Big(\sqrt{-g}\Lie_{\zeta}\varphi\p^{\mu}\varphi \Big)  = \Lie_{\zeta}\Theta^{\mu} (\delta \varphi) +  \sqrt{-g}\nabla_{\nu}({\bf A}^{\mu\nu}_{\varphi} -2{\bf S}^{\mu\nu}_{\varphi}) + \delta \varphi (...) + \delta g_{\alpha\beta}(...) \,,
\end{equation}
one obtains
\begin{equation} \label{}
{\bf A}^{\mu\nu}_{\varphi} =  0\,.
\end{equation}
This shows that the absence of the scalar contribution to the computation for the asymptotic Killing vector in our case.




\begin{thebibliography}{99} 

\bibitem{Maldacena:1997re} 
  J.~M.~Maldacena,
  ``The Large N limit of superconformal field theories and supergravity,''
  Int.\ J.\ Theor.\ Phys.\  {\bf 38}, 1113 (1999)
  [Adv.\ Theor.\ Math.\ Phys.\  {\bf 2}, 231 (1998)]
  [hep-th/9711200].


\bibitem{Henneaux:2002wm} 
  M.~Henneaux, C.~Martinez, R.~Troncoso and J.~Zanelli,
  ``Black holes and asymptotics of 2+1 gravity coupled to a scalar field,''
  Phys.\ Rev.\ D {\bf 65}, 104007 (2002)
  [hep-th/0201170].


\bibitem{Henneaux:2004zi} 
  M.~Henneaux, C.~Martinez, R.~Troncoso and J.~Zanelli,
  ``Asymptotically anti-de Sitter spacetimes and scalar fields with a logarithmic branch,''
  Phys.\ Rev.\ D {\bf 70}, 044034 (2004)
  [hep-th/0404236].


\bibitem{Bekenstein:1973ur} 
  J.~D.~Bekenstein,
  ``Black holes and entropy,''
  Phys.\ Rev.\ D {\bf 7}, 2333 (1973).


\bibitem{Hawking:1974sw} 
  S.~W.~Hawking,
  ``Particle Creation by Black Holes,''
  Commun.\ Math.\ Phys.\  {\bf 43}, 199 (1975)
  [Commun.\ Math.\ Phys.\  {\bf 46}, 206 (1976)].


\bibitem{Smarr:1972kt} 
  L.~Smarr,
  ``Mass formula for Kerr black holes,''
  Phys.\ Rev.\ Lett.\  {\bf 30}, 71 (1973)
  [Phys.\ Rev.\ Lett.\  {\bf 30}, 521 (1973)].


\bibitem{Banados:2005hm} 
  M.~Banados and S.~Theisen,
  ``Scale invariant hairy black holes,''
  Phys.\ Rev.\ D {\bf 72}, 064019 (2005)
  [hep-th/0506025].


\bibitem{Hyun:2015tia} 
  S.~Hyun, J.~Jeong, S.-A.~Park and S.-H.~Yi,
  ``Scaling symmetry and scalar hairy Lifshitz black holes,''
  arXiv:1507.03574 [hep-th].


\bibitem{Banados:1992wn} 
  M.~Banados, C.~Teitelboim and J.~Zanelli,
  ``The Black hole in three-dimensional space-time,''
  Phys.\ Rev.\ Lett.\  {\bf 69}, 1849 (1992)
  [hep-th/9204099].


\bibitem{Abbott:1982jh} 
  L.~F.~Abbott and S.~Deser,
  ``Charge Definition in Nonabelian Gauge Theories,''
  Phys.\ Lett.\ B {\bf 116}, 259 (1982).


\bibitem{Deser:2002rt} 
  S.~Deser and B.~Tekin,
  ``Gravitational energy in quadratic curvature gravities,''
  Phys.\ Rev.\ Lett.\  {\bf 89}, 101101 (2002)
  [hep-th/0205318].


\bibitem{Deser:2002jk} 
  S.~Deser and B.~Tekin,
  ``Energy in generic higher curvature gravity theories,''
  Phys.\ Rev.\ D {\bf 67}, 084009 (2003)
  [hep-th/0212292].


\bibitem{Kim:2013zha} 
  W.~Kim, S.~Kulkarni and S.-H.~Yi,
  ``Quasilocal Conserved Charges in a Covariant Theory of Gravity,''
  Phys.\ Rev.\ Lett.\  {\bf 111}, no. 8, 081101 (2013)
  [Phys.\ Rev.\ Lett.\  {\bf 112}, no. 7, 079902 (2014)]
  [arXiv:1306.2138 [hep-th]].


\bibitem{Wald:1993nt} 
  R.~M.~Wald,
  ``Black hole entropy is the Noether charge,''
  Phys.\ Rev.\ D {\bf 48}, 3427 (1993)
  [gr-qc/9307038].


\bibitem{Iyer:1994ys} 
  V.~Iyer and R.~M.~Wald,
  ``Some properties of Noether charge and a proposal for dynamical black hole entropy,''
  Phys.\ Rev.\ D {\bf 50}, 846 (1994)
  [gr-qc/9403028].


\bibitem{Wald:1999wa} 
  R.~M.~Wald and A.~Zoupas,
  ``A General definition of 'conserved quantities' in general relativity and other theories of gravity,''
  Phys.\ Rev.\ D {\bf 61}, 084027 (2000)
  [gr-qc/9911095].


\bibitem{Kim:2013cor} 
  W.~Kim, S.~Kulkarni and S.-H.~Yi,
  ``Quasilocal conserved charges in the presence of a gravitational Chern-Simons term,''
  Phys.\ Rev.\ D {\bf 88}, no. 12, 124004 (2013)
  [arXiv:1310.1739 [hep-th]].


\bibitem{Hyun:2014sha} 
  S.~Hyun, J.~Jeong, S.-A.~Park and S.-H.~Yi,
  ``Quasilocal conserved charges and holography,''
  Phys.\ Rev.\ D {\bf 90}, no. 10, 104016 (2014)
  [arXiv:1406.7101 [hep-th]].


\bibitem{Breitenlohner:1982jf} 
  P.~Breitenlohner and D.~Z.~Freedman,
  ``Stability in Gauged Extended Supergravity,''
  Annals Phys.\  {\bf 144}, 249 (1982).


\bibitem{Hertog:2004dr} 
  T.~Hertog and K.~Maeda,
  ``Black holes with scalar hair and asymptotics in N = 8 supergravity,''
  JHEP {\bf 0407}, 051 (2004)
  [hep-th/0404261].


\bibitem{Hertog:2004bb} 
  T.~Hertog and K.~Maeda,
  ``Stability and thermodynamics of AdS black holes with scalar hair,''
  Phys.\ Rev.\ D {\bf 71}, 024001 (2005)
  [hep-th/0409314].


\bibitem{Hyun:2014kfa} 
  S.~Hyun, S.-A.~Park and S.-H.~Yi,
  ``Quasi-local charges and asymptotic symmetry generators,''
  JHEP {\bf 1406}, 151 (2014)
  [arXiv:1403.2196 [hep-th]].


\bibitem{Julia:1998ys} 
  B.~Julia and S.~Silva,
  ``Currents and superpotentials in classical gauge invariant theories. 1. Local results with applications to perfect fluids and general relativity,''
  Class.\ Quant.\ Grav.\  {\bf 15}, 2173 (1998)
  [gr-qc/9804029].

\bibitem{Lu:2013ura} 
  H.~Lu, Y.~Pang and C.~N.~Pope,
  ``AdS Dyonic Black Hole and its Thermodynamics,''
  JHEP {\bf 1311}, 033 (2013)
  [arXiv:1307.6243 [hep-th]].

\bibitem{Lu:2014maa} 
  H.~Lu, C.~N.~Pope and Q.~Wen,
  ``Thermodynamics of AdS Black Holes in Einstein-Scalar Gravity,''
  JHEP {\bf 1503}, 165 (2015)
  [arXiv:1408.1514 [hep-th]].

\bibitem{Wen:2015uma} 
  Q.~Wen,
  ``Definition of Mass for Asymptotically AdS space-times for Gravities Coupled to Matter Fields,''
  arXiv:1503.06003 [hep-th].

\bibitem{Hertog:2004ns} 
  T.~Hertog and G.~T.~Horowitz,
  ``Designer gravity and field theory effective potentials,''
  Phys.\ Rev.\ Lett.\  {\bf 94}, 221301 (2005)
  [hep-th/0412169].


\bibitem{Correa:2010hf} 
  F.~Correa, C.~Martinez and R.~Troncoso,
  ``Scalar solitons and the microscopic entropy of hairy black holes in three dimensions,''
  JHEP {\bf 1101}, 034 (2011)
  [arXiv:1010.1259 [hep-th]].
  
  
  
\bibitem{Iizuka:2015vsa} 
  N.~Iizuka, A.~Ishibashi and K.~Maeda,
  ``A rotating hairy AdS$_3$ black hole with the metric having only one Killing vector field,''
  arXiv:1505.00394 [hep-th].

\bibitem{Hertog:2006rr} 
  T.~Hertog,
  ``Towards a Novel no-hair Theorem for Black Holes,''
  Phys.\ Rev.\ D {\bf 74}, 084008 (2006)
  [gr-qc/0608075].

\bibitem{Faulkner:2010fh} 
  T.~Faulkner, G.~T.~Horowitz and M.~M.~Roberts,
  ``New stability results for Einstein scalar gravity,''
  Class.\ Quant.\ Grav.\  {\bf 27}, 205007 (2010)
  [arXiv:1006.2387 [hep-th]].






\bibitem{Gonzalez:2011nz} 
  H.~A.~Gonzalez, D.~Tempo and R.~Troncoso,
  ``Field theories with anisotropic scaling in 2D, solitons and the microscopic entropy of asymptotically Lifshitz black holes,''
  JHEP {\bf 1111}, 066 (2011)
  [arXiv:1107.3647 [hep-th]].


\end{thebibliography}
\end{document}